\documentclass[graybox]{svmult}

\usepackage{mathptmx}       
\usepackage{helvet}         
\usepackage{courier}        
\usepackage{type1cm}        
\usepackage{makeidx}         
\usepackage{graphicx}        
\usepackage{multicol}        
\usepackage[bottom]{footmisc}

\usepackage{amsmath}
\usepackage{amssymb}
\usepackage{amsfonts}

\makeindex             

\begin{document}

\title*{Time allocation in social networks: correlation between social structure and human communication dynamics}
\titlerunning{Correlation between social structure and human communication dynamics}
\author{Giovanna Miritello, Rub\'en Lara, and Esteban Moro}
\institute{Giovanna Miritello, Departamento de Matem\'aticas \& GISC, Universidad Carlos III de Madrid, 28911 Legan\'es, Spain \& Telef\'onica R\&D, Madrid, Spain, 
\and Rub\'en Lara, Telef\'onica R\&D, Madrid, Spain 
\and Esteban Moro, Departamento de Matem\'aticas \& GISC, Universidad Carlos III de Madrid, 28911 Legan\'es, Spain \& Instituto de Ciencias Matem\'aticas {CSIC-UAM-UCM-UC3M}, 28049 Madrid, Spain \& Instituto de Ingenier\'{\i}a del Conocimiento, Universidad Aut\'onoma de Madrid, 28049 Madrid, Spain \email{emoro@math.uc3m.es}}
%
%
\maketitle

\abstract{Recent research has shown the deep impact of the dynamics of human interactions (or temporal social networks) on the spreading of information, opinion formation, etc. In general, the bursty nature of human interactions lowers the interaction between people to the extent that both the speed and reach of information diffusion are diminished. Using a large database of 20 million users of mobile phone calls we show evidence this effect is not homogeneous in the social network but in fact, there is a large correlation between this effect and the social topological structure around a given individual. In particular, we show that social relations of hubs in a network are relatively weaker from the dynamical point than those that are poorer connected in the information diffusion process. Our results show the importance of the temporal patterns of communication when analyzing and modeling dynamical process on social networks.}

\section{Introduction}\label{sec:zero}

A quantitative understanding of human communication patterns is of paramount importance not only for a better understanding of human behavior, but also to explain the dynamics of many social, technological and economic phenomena. Examples include epidemics spreading, virus outbreaks, opinion formation, diffusion of innovation, rumors or trends \cite{lazer2009,pastorsatorras2001a,rogers1995}.
All these processes are related to the underlying structure of the network and on the temporal activity patterns of humans, since they depend on the way humans interact and share information \cite{barabasi2005,iribarren2009,malmgren2008}. Determination of spreading paths, speed and reach in society is crucial for developing efficient strategies to propagate information (like in viral marketing \cite{iribarren2009}), to find out the influence that different people play in social networks \cite{watts2007} or to know how public opinion forms and spreads \cite{gonzalezbailon2011}. 

Most of our current understanding of spreading phenomena comes from implementing models and ideas borrowed from epidemiology on empirical or synthetic social networks \cite{anderson1992}. Although recent works suggest that information transmission or influence is a much more involved process than disease propagation \cite{aral2012,iribarren2011b,ugander2012}, these studies have permitted to find the deep entanglement between spreading and the complex topological patterns of the underlying network in the dynamical process \cite{barrat2008,barthelemy2004,boguna2002,kitsak2010,onnela2011}. 
Social networks are very heterogeneous: the flux of information that pass through each social tie is unevenly distributed, some individuals are more connected than others, social relationships are organized into communities, etc. These network structure heterogeneities reflect the way in which whom and how each individual is connected (and located) within the network and affect, therefore, the spreading of information. 
In particular, a celebrated hypothesis is that "hubs" (most connected people) \cite{pastorsatorras2001a} or those users with large centrality \cite{kitsak2010} are the key players in diffusion processes, being responsible for the largest part of the total reach of the spreading process.

Paradoxically, most of these studies of dynamical phenomena on social networks neglect the {\em temporal patterns} of human communications: humans act in bursts or cascades of events \cite{barabasi2005,isella2011,rybski2009,vazquez2007}, most social ties are not persistent \cite{hidalgo2008,kossinets2006} and communications happen mostly in the form of group conversations \cite{eckmann2004,isella2011,wu2010,zhao2010}. Since information transmission and human communication are concurrent, the temporal structure of communication must influence the properties of information spreading. Indeed, recent experiments of electronic recommendation forwarding \cite{iribarren2009} and simulations of epidemic models on email and mobile databases \cite{karsai2011,miritello2011,vazquez2007} found that the asymptotic speed of information spreading is controlled by the bursty nature of human communications that leads to a slowing down of the diffusion. Furthermore, in \cite{miritello2011} it was found that not only the speed, but also the reach of information is affected by the temporal patterns of human communication. Specifically, while {\em burstiness of human communication} hinders the propagation of information through social ties, {\em group conversations} (or correlation between communication events in neighbor ties) favor the formation of local information cascades. These two competing effects shape the spreading reach and speed of information yielding to different possible behaviors depending on the temporal properties of the information transmission. 

Although identification of the key ingredients of the dynamical patterns of human communication allows to understand the evolution of the process at the network level, still there is no clear knowledge of what the roles played by the individual patterns of activity in that process are and how they correlated with the (static or aggregated) topological structure of social networks. For example, we might wonder whether static hubs, identified as those who contact a large number of people in a given time period, are also dynamical hubs, i.e., they have such communication patterns so that they retain their key role in information diffusion on temporal networks. In this contribution we analyzed this hypothesis by investigating how the different properties of temporal patterns of communication correlate with the social connectivity of an individual. More generally, this problem is related to how people allocate time among their social connections. After all, time and attention are inelastic resources and thus humans have to implement a communication strategy to maintain their social connectivity. Although static models of information diffusion assume that all ties of a given node are open at all times and information can flow at any time through any given tie, we expect attention and time constrains to shape the dynamical properties of human communication. Analyzing the communication of mobile phone calls between people, our results suggest that ties of static hubs have less transmissibility (on average) than expected and thus we find a large correlation between the static (node's degree) and dynamical properties of human communication. Taken together, these results suggest that temporal patterns of communication must be incorporated in the description and modeling of dynamical human-driven phenomena and of quantitative models of contact social networks.

This contribution is organized as follows: in Section \ref{sec:one} we describe our data and methods and review the different ways to characterize communication patterns between people. In Section \ref{sec:two} we revisit our method introduced in \cite{miritello2011} in which we map the dynamic of human interactions onto a static representation of a social network through a quantity we call the {\em dynamical strength of ties}. We also study the correlation between the dynamical patterns and social connectivity of human communication in that section. Finally in the discussion we comment on our results and the possibility of using the dynamical strength of ties as a way to model temporal networks.

\section{Characterizing human communication patterns}\label{sec:one}
\begin{figure}[t]
\begin{center}
\includegraphics[width=1\textwidth,clip=]{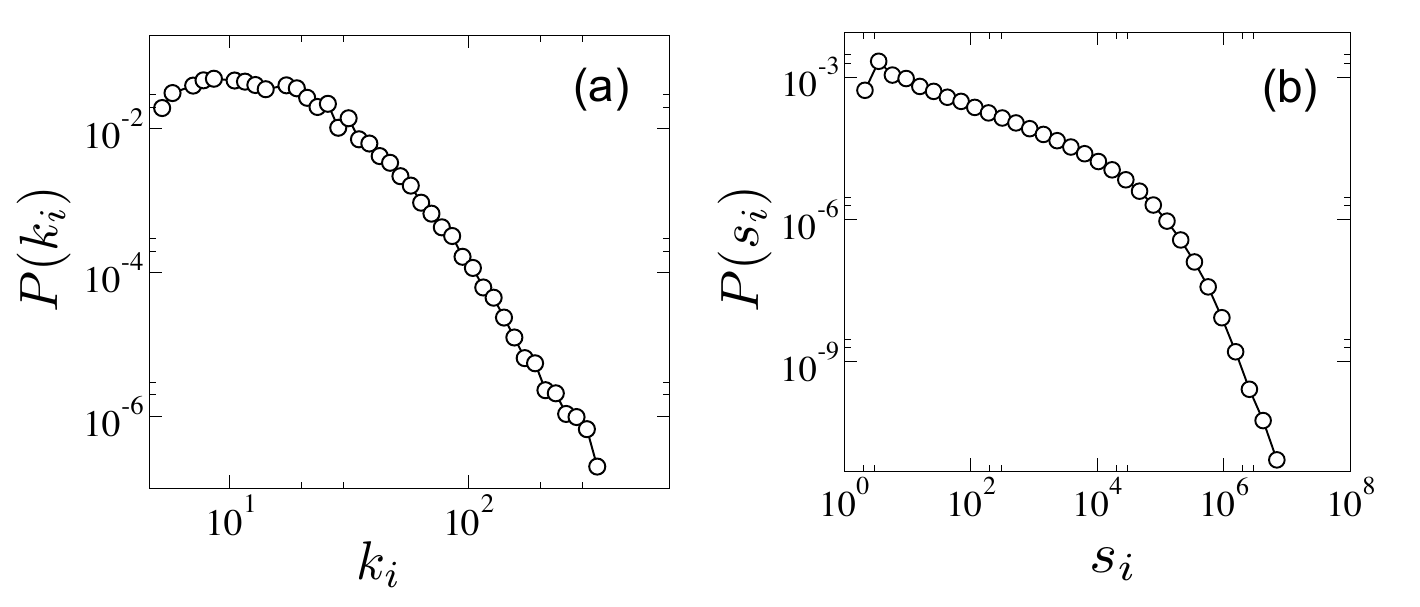}
\end{center}
\caption{Distribution of the social connectivity $k_i$ (a) and the nodes' strenght (b) in our mobile phone network.}
\label{fig:pdfks}
\end{figure}

To understand what features characterize human communication patterns, we study the mobile phone calls during $T_0 = 11$ months of a single mobile phone operator in a given country. The data consists of $2 \times 10^7$ phone numbers and $7 \times 10^8$ communication ties for a total of 9 billion calls between users. Call Detail Record (CDR) contains the hashed number of the caller and the receiver, the time when the call was initiated and the duration of the call. We consider only events in which the caller and the callee belong to the operator under consideration, because of partial access to the records of other operators. Finally, we only consider ties which are reciprocated, i.e., in which a call is made at least in both directions $i \leftrightarrow j$ and thus we consider that the weight of the tie $w_{ij}$ given by the number of calls or the total amount of call time between $i$ and $j$, is symmetric ($w_{ij} = w_{ji}$).\footnote{Similar results are found if the total amount of time is used for $w_{ij}$ instead of the number of calls.} Our data for the connectivity of the social network, the duration of the calls, etc. are very similar to those reported in previous studies of mobile phone networks \cite{onnela2009}. As shown in Fig.~\ref{fig:pdfks}, we found a skewed distribution for both the nodes' social connectivity $k_i$ in the observation time period and the strength $s_i = \sum_j w_{ij}$.
In our database, we observe that the mean social connectivity is around 85, with a maximum value of around 500.
For the node strength $s_i$ instead, we found that, although the mean of this distribution is around 1.5 hours in the whole period, the maximum value is about 6 hours per day. This means that while the time that the larger part of the population spends on the phone per day is of the order of seconds or minutes, there is a small minority who phone more than 1 hour per day. Not only the aggregated $s_i$, but also the ties weight $w_{ij}$ show a long-tailed distribution, which indicates a strong heterogeneity in the way people distribute the time across their social circle. For both $k_i$ and $s_i$, however, the decay is faster than a power-law, which indicates the presence of a relatively small number of hubs. This decay is probably due to the fact that we have filtered out business phone numbers, which mainly correspond to the hubs in mobile networks, a possibility also pointed out in \cite{onnela2007a}.

The relationship between the social connectivity and the intensity of communication has been investigated recently by Miritello {\em et al.} in \cite{miritello2012b}: in line with previous studies for both scientific collaboration and the air-transportation networks \cite{barrat2004}, we found that the average strength $s_i(k_i)$ of nodes with degree $k_i$ increases almost linearly with $k_i$, that is $s_i(k_i) \sim k_i^\beta$, with $\beta\simeq 1$. However, we observe a slightly more complex behavior for large values of $k_i$, where $s_i$ starts to grow sublinearly with $k_i$ until it saturates for very large values of $k_i$, which suggests the existence of a limit for more connected people to allocate communication time in a proportional fashion. 
Thus, the larger the social connectivity, the smaller the time dedicated per tie. This means that, on average, hubs have weaker communications. The observed behavior might be related to Dunbar's theory \cite{dunbar1998} which asserts that cognitive and biological constraints limit the number of people an individual keeps social contact with. On top of those limits, in our case also temporal and monetary constraints may play their role in phone communication. In this respect, our results are similar to those found in other communication networks such as Twitter \cite{goncalves2011}.

Characterizing a given communication tie by its strength (weight $w_{ij}$) ignores however the fact that human communication is a highly complex dynamical process. Among others, some of the implicit assumptions of considering static description of ties are that (i) communication can happen at any time, (ii) human activity is Markovian and randomly distributed in time, therefore well approximated by a Poisson process, and (iii) there is no correlation or causality between events.

\begin{figure}[t]
\begin{center}
\includegraphics[width=1\textwidth,clip=]{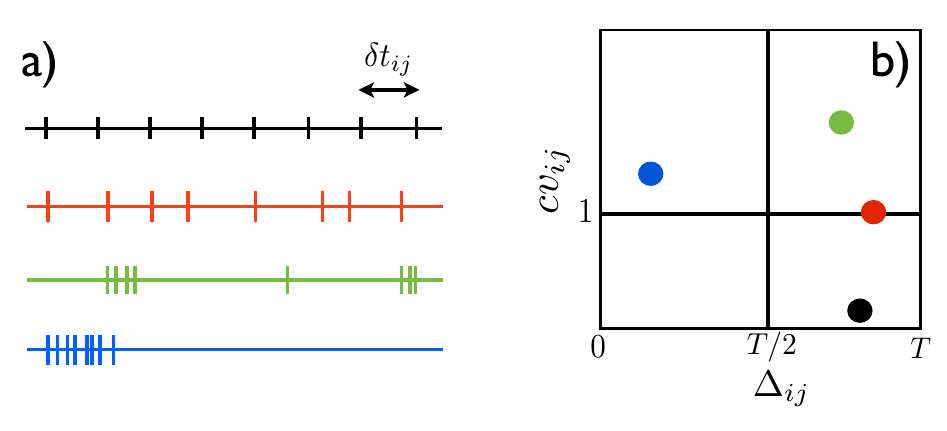}
\end{center}
\caption{(a) Schematic representation of different communication patterns within a tie for the same intensity, $w_{ij} = 8$ calls. Each vertical line correspond to a different call between $i$ and $j$. (b) Bivariate representation of the coefficient of variation and stability of ties for the corresponding cases in (a).}
\label{fig:scheme}
\end{figure}

This is exemplified in Fig.~\ref{fig:scheme}, where we show that many different temporal patterns of communication can correspond to the same strength of communication $w_{ij}$. In particular, recent research has shown that temporal patters of human individuals are strongly inhomogeneous and deviate from the Poisson process \cite{barabasi2005,karsai2011,miritello2011,vazquez2007}. In the latter, the number of events during a time interval of duration $T$ follows a Poisson distribution with mean $\rho T$, where $\rho$ is the rate of events and the time between consecutive events, called the {\em inter-event} times, follows an exponential distribution $p(\delta t) = \rho e^{-\rho t}$. As a consequence, individual actions happen at relatively regular time intervals $\delta t$ and very short or very long inter-event times occur with small probability. However, human activity is bursty, which is reflected by the slowly decaying of the inter-event time probability distribution (possibly like $P(\delta t) \sim \delta t^{-1}$ for small inter-event times \cite{barabasi2005}), thus in stark contrast with the prediction of a homogeneous Poisson process. This behavior seems to be a universal feature of human activity: it has  been observed in several systems driven by human activity sequences \cite{barabasi2010,eckmann2004,goh2008,oliveira2005,rybski2010} and is known in literature as {\it bursty behavior} since long periods of inactivity are separated by intense bursts of activity. Regarding human communication, bursty behavior is found also in inter-event times between communications of a single person and even on events within a social tie \cite{karsai2011,miritello2011}. Although there are other ways to characterize (and define) burstiness \cite{goh2008,karsai2012}, to capture this information we use the common {\em coefficient of variation} $cv_{ij}$ of the inter-event time distribution $P(\delta t_{ij})$ of a tie. It is measured as  $cv_{ij} = \sigma_{ij}/\mu_{ij}$ where $\mu_{ij}$ and $\sigma_{ij}$ are the mean and standard deviation of the distribution $P(\delta t_{ij})$.  By definition, $cv_{ij}$ measures the dispersion of a distribution, that is the level of heterogeneity of tie communication: the more heterogeneous the interaction is, the larger $cv_{ij}$.
Note that for the inter-event exponential distribution in the Poisson process we have that $\sigma_{ij} = \mu_{ij}$ and thus $cv_{ij} = 1$. For the perfect deterministic process in which events happen at regular times we get $cv_{ij} = 0$, while for the bursty behavior in which $P(\delta t)$ is heavy tailed we usually have $\sigma_{ij} \gg \mu_{ij}$ and thus $cv_{ij} \gg 1$.

Although $w_{ij}$ and $cv_{ij}$ tell us something about the intensity and how homogeneous in time are communication events, they lack the information about how fast this happens. As shown in Fig.~\ref{fig:scheme}, the same amount of calls can be place in a rather long or extremely short time window. To characterize this variability in the rhythm of communication we define the temporal stability $\Delta_{ij}$ of a tie as $\Delta_{ij} = t_{ij}^{max} - t_{ij}^{min}$, where $t_{ij}^{min}$ and $t_{ij}^{max}$ are, respectively, the time instants at which the first and the last communication events between $i$ and $j$ are observed within the observation period. A large stability ($\Delta_{ij} \simeq T_0$) indicates that the communication between $i$ and $j$ extends over the observation period, while a small one ($\Delta_{ij} \simeq 0$) is the signal of a short tie lifetime. In the case of a Poissonian process $\Delta_{ij} \simeq T_0$ for all ties, since time events are evenly distributed along the observation window.

\begin{figure}[t]
\begin{center}
\includegraphics[width=1\textwidth,clip=]{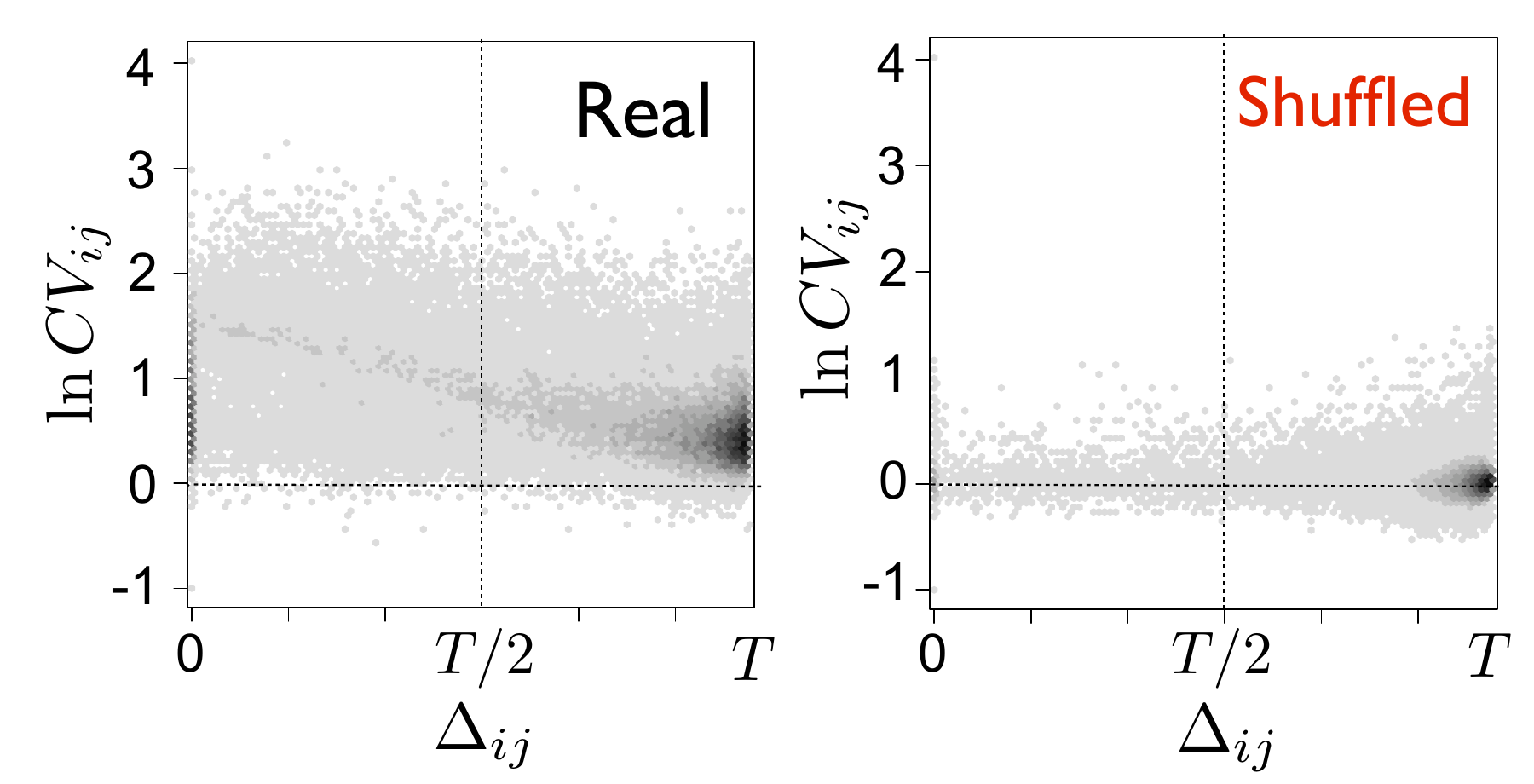}
\end{center}
\caption{Density plot of the coefficient of variation and stability of the ties in our database for the real sequence of calls (left) and the corresponding ones for the shuffled time stamps (right). To have enough statistical significance for $cv_{ij}$, only ties with $w_{ij} \geq 5$ are used in the plot.}
\label{fig:scheme2}
\end{figure}

As shown in Fig.~\ref{fig:scheme2}, human communication patterns differ from equivalent Poisson process with the same number of events $w_{ij}$. To mimic the latter process, we shuffle the time stamps of the real events across the database,  thus each call has an even probability to appear anytime within the observation window. Note that this shuffling preserves the usual circadian rhythms (nights, weekends, holidays), but destroys all possible heterogeneous patterns of communication within the ties \cite{jo2012}. As expected, in the shuffled Poissonian case we have that for most of the ties $cv_{ij} \simeq 1$ and $\Delta_{ij} \simeq T_0$. However, for the real case we observe that most ties show bursty behavior ($cv_{ij} > 1$) and that the stability is distributed along the observation window, with the special feature that a large proportion of ties have either large stability ($\Delta_{ij} \simeq T_0$) or very small one ($\Delta_{ij} = 1$). This  bimodal distribution of the lifetime of links was found also in other mobile phone databases \cite{hidalgo2008} and indicates that ties are mostly long or very short, possibly signaling the very different nature of the communication involved: for example, short ties could be due to search of information within the social network, while long ties might reflect social or close relationship between individuals.

\begin{figure}[t]
\begin{center}
\includegraphics[width=1\textwidth,clip=]{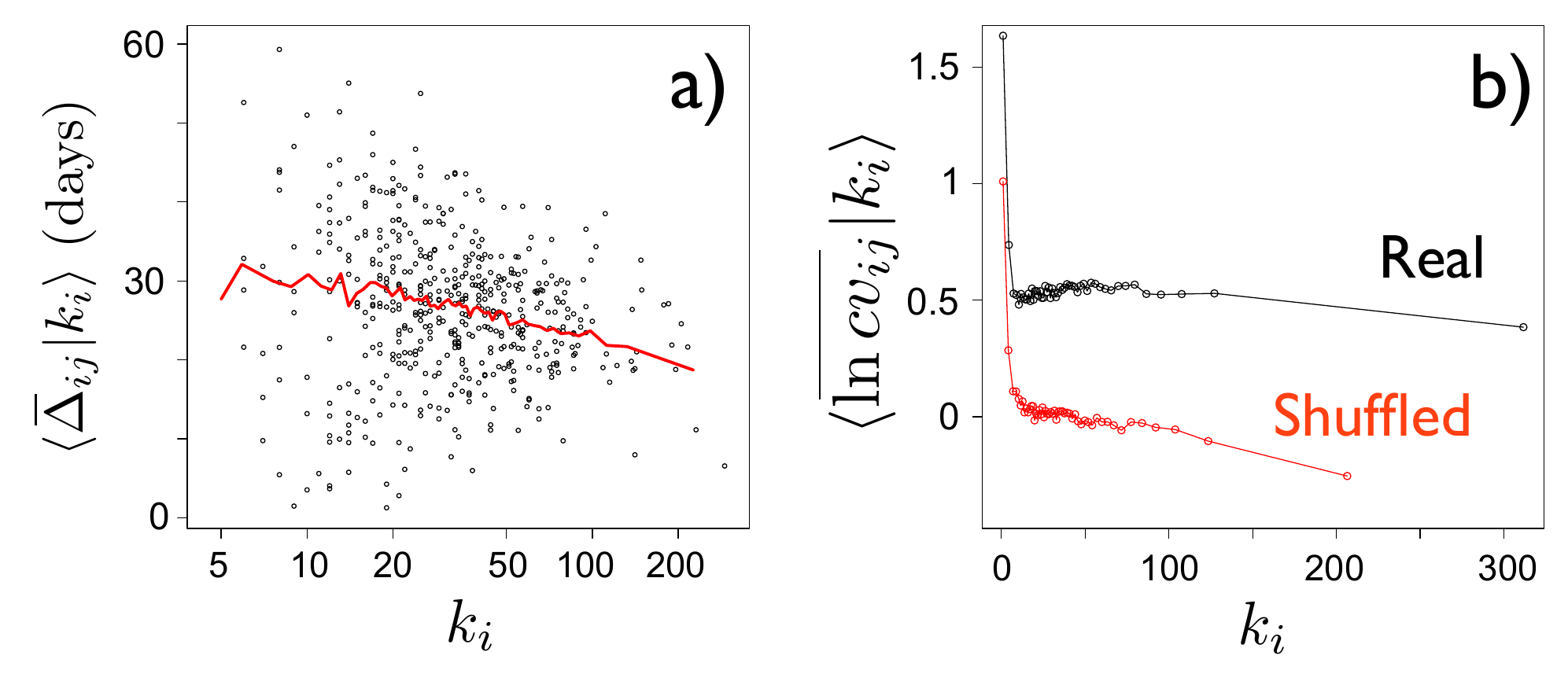}
\end{center}
\caption{(a) Scatter plot (for a small sample of individuals) and conditional mean of the average stability of the ties for a given individual $\overline{\Delta}_{ij}$ in our database as a function of her connectivity in the observation window $k_i$. The plot only shows the result for the real sequence of calls, since for the shuffled case $\overline{\Delta}_{ij} \simeq T$. (b) Conditional mean of the average logarithm of the coefficient of variation $cv_{ij}$ for the real and shuffled cases as a function of $k_i$.}
\label{fig:dki}
\end{figure}

We now investigate how the dynamical properties of ties depend on the social connectivity of individuals in the observation time window $k_i$. 
As mentioned above, cognitive and monetary costs influence the aggregate amount of attention per tie for largely connected individuals \cite{dunbar1998,goncalves2011,miritello2012b}.
On the other hand, we have seen that communication events are not evenly distributed across the time window. We therefore expect that the way in which people allocate time in social networks also reflects in their patterns of communication.
For example, for a given $w_{ij}$ the attention allocated in a short tie is much more localized in time than in a long tie. In addition, individuals might choose to develop more bursty communication patterns to be able to allocate more conversations within the day. 
In Fig.~\ref{fig:dki} we show our results for the dependence of both $cv_{ij}$ and $\Delta_{ij}$ with $k_i$. Although there is no significant dependence on $cv_{ij}$, i.e., the burstiness of ties, for a given individual as a function of her social connectivity $k_i$ in the time window, we observe that highly connected individuals have shorter ties. 
This is a clear indication that, aside from the non-trivial way in which individuals allocate time of communication among their contacts, there is also a complex way in which this attention unfolds in time. In summary, it seems that highly connected people are characterized by weaker (in terms of volume of communication) and shorter ties than moderate or low connected people.

\section{Dynamical strength of communication ties}\label{sec:two}
It is clear that the intensity $w_{ij}$ does not capture the whole dynamics of communication within a tie and thus it is not sufficient to describe the importance or role of that tie in the diffusion of information. To understand that role, in \cite{miritello2011} we developed a new measure of the strength of a tie which takes into account not only the intensity of the link, but also its dynamical pattern. The idea is based on the mapping introduced by Newman in \cite{newman2002a} in which the dynamical information diffusion process is mapped onto a static edge percolation where each tie is described by the {\em transmissibility} or {dynamical strength}, i.e., the probability that information flows through the link given the sequence of communication events between individuals. The network is then still described by a static graph, but the interaction strength between individuals now incorporates the causal and temporal patterns of their communications. As shown in \cite{miritello2011}, this procedure not only explains the qualitative behavior of the dynamics of information diffusion, but also successfully predicts, for example, the percolation threshold for the SIR (Susceptible-Infected-Recovered) model on our dataset.
Obviously the dynamical strength of a tie must depend on the process under consideration. For example, its definition varies if one considers models of information diffusion like the SIR  model \cite{miritello2011}, simple random walk hopping between individuals \cite{hoffman2011}, or more complicated processes. In our case, since we are interested on information diffusion and/or influence, we concentrate on the simplest model of information propagation, namely the SIR epidemiological model \cite{anderson1992}. Although recent research has highlighted that important differences exist between information diffusion and disease spreading \cite{aral2012,iribarren2011b,ugander2012}, epidemiological models have been the main theoretical tools to understand how information is transmitted in social networks \cite{daley1964}. Moreover, they are useful methods to explore the intricate structure of social networks at large scale \cite{karsai2011,kitsak2010,onnela2011}. In that spirit, our aim in this section is to use the SIR model to investigate the influence of the behavior observed in the previous section on the role played by individuals in the spreading of information.

In the SIR model, individuals can be at different states of the infection dynamics and they are allowed to change their state. Individuals are initially susceptible (S) and become infected (I) with a given transmission rate $\lambda$ when interact with an infected individual. At the same time, infected individuals are allowed to recover (R) at some rate $\mu$. In our case, we consider that in each call an infected node (an individual that knows the information) can infect a susceptible node (an individual who is not aware of the information) with probability $\lambda$. Due to the synchronous nature of the phone communication, this happens regardless of who initiates the call. Nodes remain infected for a certain interval of time until they decay into the recovered state in which they do not propagate the information anymore. Although this decay into the recovered state can be very complex, for the sake of simplicity we simulate the simplest model in which nodes recover after a fixed deterministic and homogeneous time $T$. More general situations can be accommodated into the model such as, for example, stochastic recovery times. The information diffusion is simulated starting from a unique infected seed which generates a viral cascade that grows until there are no more nodes in the infected state. As shown in \cite{miritello2011}, the simulations using the real sequence of calls between individuals show a phase transition at very small $\lambda_c$: below $\lambda_c$ cascades die out very quickly, while above $\lambda_c$ we find that with large probability there exists a cascade that infects a large proportion of the individuals in our social network. This percolation transition happens also around $\lambda_c$ for the  time-shuffled data, in which the real sequence and properties of tie interactions are destroyed mimicking Poissonian dynamics. 
A different behavior was however found between the real and the shuffled case below and above the percolation transition $\lambda_c$. Below $\lambda_c$, the correlations between events in neighboring ties (group conversations) favor information diffusion and thus cascades in the real case are larger than in the shuffled case. In contrast, above the percolation threshold, the burstiness of communication events hinders the propagation of information, making the reach of information smaller in the real than in the shuffled case. 
Thus real dynamics of interaction in social network makes information spreading more efficient at small (local) scale (below $\lambda_c$), while if information propagates easily (large $\lambda$) the reach of informaction in social networks is small than the one expected when a Poissonian dynamics is considered. 

\begin{figure}[t]
\begin{center}
\includegraphics[width=0.8\textwidth,clip=]{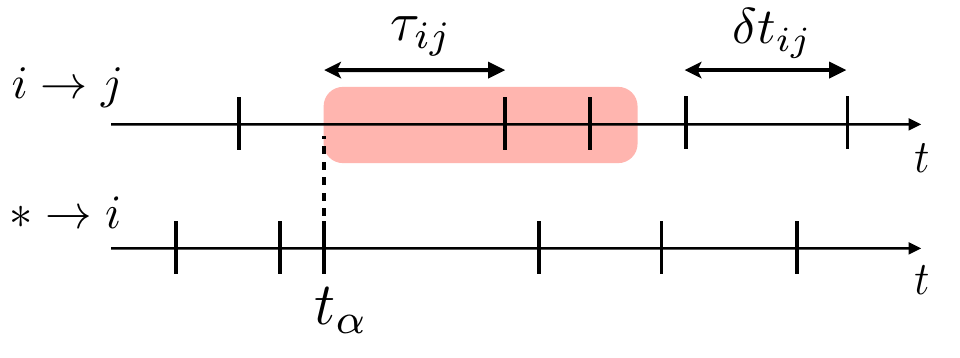}
\end{center}
\caption{\textit{Schematic view of communications events around individual $i$: each vertical segment indicates an event between $i\to j$ (top) and $*\to i$ (bottom). At each $t_{\alpha}$ in the $* \to i$ time series, $\tau_{ij}$ is the time elapsed to the next $i \to j$ event, which is different from the inter-event time $\delta t_{ij}$ in the $i \to j$ time series. The red shaded area represents the recover time window $T$ after $t_{\alpha}$.}}
\label{fig:arrows}
\end{figure}

The observed behavior can be understand by analyzing how the communication temporal patterns affect information diffusion \cite{miritello2011}. Spreading from user $i$ to user $j$ ($i \to j$) happens at the {\em relay time intervals} $\tau_{ij}$, i.e., the time interval it takes to $i$ to pass on to $j$ an information he/she got from any another person $* \to i$, where $j\neq *$ (see Fig.~\ref{fig:arrows}). 
Information spreading is thus determined by the interplay between $\tau_{ij}$ and the intrinsic timescale of the infection process $T$. 
As it was shown in \cite{miritello2011}, $\tau_{ij}$ depends on the correlated and causal way in which group conversations happen, since it depends on the inter-event intervals $\delta t_{ij}$ in the $i \to j$ communication but {\em also} on the possible temporal correlation with the $* \to j$ events \cite{kenan2007,newman2002a}.
\\By ignoring this correlation the probability distribution function (or pdf) for $\tau_{ij}$ can be approximated by the waiting-time density for $\delta t_{ij}$ given \cite{breuer2005}:
\begin{equation}
P(\tau_{ij}) = \frac{1}{\overline{\delta t}_{ij}} \int_{\tau_{ij}}^\infty P(\delta t_{ij}) ~d \delta t_{ij},
\label{waitingtime}
\end{equation}
where $\overline{\delta t}_{ij}$ is the average inter-event time. In this approximation, the dynamics of the transmission process only depends on the dyadic $i\to j$ sequence of communication events. In particular, the heavy-tail properties of $P(\delta t_{ij})$ found in human communication \cite{karsai2011,miritello2011} are directly inherited by $P(\tau_{ij})$, making the relay times much bigger than expected. Thus the more bursty the communication is, the larger the relay times are, which qualitatively explains empirical observations that heterogeneous human activity slows down information spreading \cite{iribarren2009,iribarren2011,karsai2011,vazquez2007}.

To fully understand the impact of dynamical patterns of communication on information diffusion, we follow the approach of \cite{newman2002a} by mapping the dynamical SIR model to a static edge percolation model where each tie is described by the {\it transmissibility} ${\cal T}_{ij}$. The transmissibility represents the probability that the information is transmitted from $i$ to $j$ and it is a function of $\lambda$ and $T$.
If user $i$ becomes infected at time $t_\alpha$ and the number of communication events $i\to j$ in the interval $[t_\alpha,t_\alpha+T]$ is $n_{ij}(t_\alpha)$, then the transmissibility in that interval is (see Fig.\ref{fig:arrows})
\begin{equation}
{\cal T}_{ij} = 1 - (1-\lambda)^{n_{ij}(t_\alpha)}.
\label{eq:tras1}
\end{equation}
User $i$ may become infected during any $* \to i$ communication event at $t_\alpha$. 
Assuming these events independent and equally probable, we can average ${\cal T}_{ij}$ over all the $t_\alpha$ events to get
\begin{equation}\label{eq:tras2}
{\cal T}_{ij}[\lambda,T] = \langle 1-(1-\lambda)^{n_{ij}(t_\alpha)}\rangle_{\alpha}.
\end{equation} 
If the number of  $* \to i$ events is large enough, we can use a probabilistic description of Eq.(\ref{eq:tras2}) in terms of the probability $P(n_{ij}=n;T)$ that the number of communication events between $i$ and $j$ in a given time interval $T$ is $n$. Thus
\begin{equation}
{\cal T}_{ij}[\lambda,T] = \sum_{n=0}^\infty P(n_{ij}=n;T) [1-(1-\lambda)^n],
\label{eq:tras3}
\end{equation}
which in principle can be non symmetric (${\cal T}_{ij} \neq {\cal T}_{ji}$).
This quantity represents the real probability of infection from $i$ to $j$ and defines what we called the {\em dynamical strength} of the tie. Note that ${\cal T}_{ij}$ depends on the series of communication events between $i$ and $j$, but also on the time series of calls received by $i$. 

In \cite{newman2002a} Newman studied the case in which both time series are given by independent Poisson processes in the whole observation interval $[0,T_0]$. 
Thus, $P(n_{ij}=n;T)$ is the Poisson distribution with rate $\rho_{ij} = w_{ij} T / T_0$, where $w_{ij}$ is the intensity of the 

\clearpage

\begin{figure}[t]
\begin{center}
\includegraphics[width=1\textwidth,clip=]{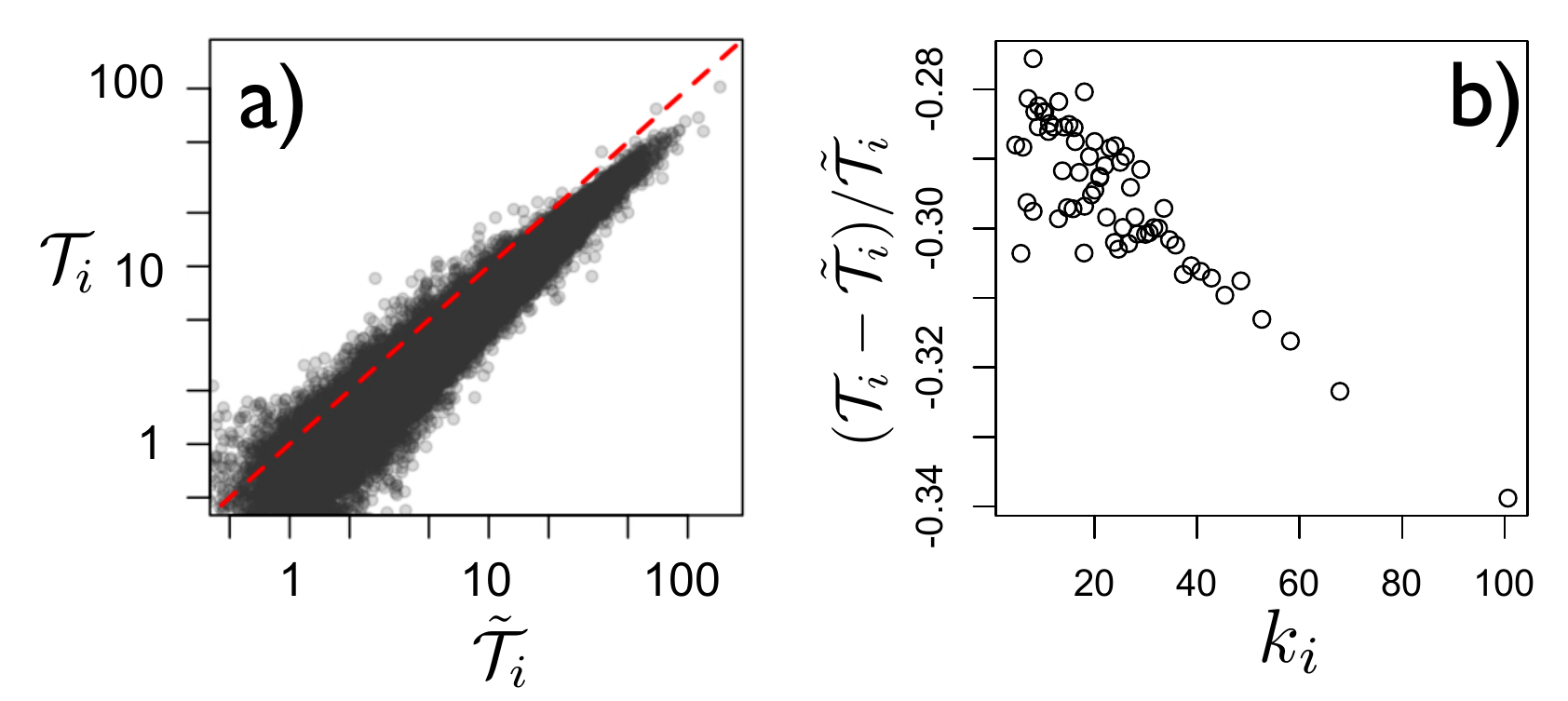}
\end{center}
\caption{(a) Comparison between the total transmissibility of individuals in the real ${\cal T}_{i}$ and shuffled $\tilde {\cal T}_i$ call records for a random set of $10^5$ users in our database. Dashed line is the ${\cal T}_{i} = \tilde {\cal T}_{i}$ line. (b) Conditional average of relative difference between the real and shuffled total transmissibility as a function of the social connectivity $k_i$.}
\label{fig:tiki}
\end{figure}

\noindent
tie, i.e., the total number of calls from $i$ to $j$ in $[0,T_0]$, and so
\begin{equation}
\tilde {\cal T}_{ij}[\lambda,T] = 1 - e^{-\lambda \rho_{ij}} = 1 - e^{-\lambda w_{ij}T /T_0},
\end{equation}
which shows the one-to-one relationship between the intensity $w_{ij}$ and the transmissibility ${\cal T}_{ij}$ in the Poissonian case: the more intense the communication is, the larger the probability of infection. 
However, as we have shown in \cite{miritello2011} and in previous sections, the real $i \to j$ and $*\to i$ series are far from being independent and Poissonian. To proceed analytically, we approximate Eq.~(\ref{eq:tras2}). 
For small values of $\lambda$ we have $1-(1-\lambda)^n \simeq \lambda n$, while when $\lambda \simeq 1$ we get that $1-(1-\lambda)^n \simeq 1$ for $n > 0$. Thus, the transmissibility for the two regimes is given by:

\begin{equation}\label{eq:tras4}
{\cal T}_{ij}[\lambda,T] = \left \{
\begin{array}{ll}
\lambda~\langle n_{ij}\rangle_{t_\alpha} & \mathrm{for}\ \lambda \ll 1\\
1 - P_{ij}^0 & \mathrm{for}\ \lambda \simeq 1
\end{array}\right.
\end{equation}

\noindent
where $P_{ij}^0 = P(n_{ij}=0;T)$ is the probability of no communication event in a time window of length $T$. 
This approximation allows us to estimate ${\cal T}_{ij}$ in a much simpler way, since it depends now only on variables that can be measured from the temporal activity. In fact, Eq.~(\ref{eq:tras4}) explains the observed behavior found in \cite{miritello2011}, since group conversations make $\langle n_{ij}\rangle_{t_\alpha}$ bigger in the real case than in the Poisson approximation, while burstiness yields to larger values of the probability of no communication in the real case than in the Poisson case and thus transmissibility is larger in the latter than in the former situation. 

For $\lambda \simeq 1$, the larger the coefficient of variation $cv_{ij}$ in a tie (i.e. the more bursty it is), the larger the probability of no event $P_{ij}^0$ and thus the lower the transmissibility of the tie. The same behavior is found for the lifetime of links: if communication is concentrated in a short time period, then $P_{ij}^0$ is large and thus the transmissibility is low. Consequently, the shorter the tie communication, the smaller the transmissibility of the ties. Since the burstiness of human communication and activity seem to be universal \cite{barabasi2005}, empirical description of social networks only by their topological structure and/or the intensity $w_{ij}$ typically overestimates the transmissibility power of ties. 

Finally, we investigate whether this effect of human communication patterns is correlated with the social structure around a particular node. As shown in the previous section, intensity and lifetime of links are distributed differently among neighbors for individuals with low and large social connectivity. Since lifetime and burstiness of communication ties impact the transmissibility of links, we expect that the dependence of time allocation of communication with connectivity also translates into a dependence between $k_i$ and the total transmissibility of an individual ${\cal T}_{i} = \sum_j {\cal T}_{ij}$. Note that in the Poisson case and for small values of $\lambda$, we get that $\tilde {\cal T}_{ij} \simeq \lambda w_{ij} T/T_0$ and thus $\tilde {\cal T}_i \simeq \lambda\; s_i\; T/T_0$, i.e., the transmission power of a node is proportional to the strength of the node in this approximation. However, in the real case ${\cal T}_i = \sum_j {\cal T}_{ij}[\lambda,T]$ depends both on the intensity and dynamical patterns of communication. As we see in Fig.~\ref{fig:tiki}a, the burstiness of ties makes individuals in general less powerful to transmit information than in the Poisson case, in the sense that their total real transmissibility is smaller than shuffled one. However, this is not an homogeneous effect in our database since, as it can be seen in Fig.~\ref{fig:tiki}b, the relative difference between ${\cal T}_i$ in the real and shuffled cases is larger for hubs or more connected people than for poorly connected people, a manifestation in the process of information diffusion of the effect we have seen in the previous section (i.e. hubs have shorter ties). Note that this relative difference increases with $k_i$ meaning that ${\cal T}_i$ does not grow linearly with $k_i$ (or $s_i$) as $\tilde {\cal T}_i$ does. Our findings show that neither $k_i$ or $s_i$ are good predictors of the local spreading power or influence of a node, especially for largely connected people or hubs. In this sense, although in general static hubs (people with large $k_i$ or $s_i$) have also large dynamical transmissibility, the large variability shown in Fig.~\ref{fig:tiki}a implies that this correspondence is not always true.

\section{Discussion}
In the last years there has been an increasing interest in characterizing the complex topological structure of the underlying contact network and in understanding how it affects the diffusion of information, innovation or opinions \cite{newman2003a}.
However, most of these studies neglect the temporal dimension of human communication as the fact that humans act in bursts or cascades of events \cite{barabasi2005,isella2011,miritello2011,karsai2011,vazquez2007,rybski2009}, most of ties form and decay within the observation time period \cite{hidalgo2008,kossinets2006} and there are correlations between communication events \cite{eckmann2004,isella2011,miritello2011,wu2010,zhao2010}.
These temporal heterogeneities significantly affect the current description of social networks, where ties are described by a static strength which usually only takes into account the volume of communication or intensity $w_{ij}$.  Although $w_{ij}$ has been shown to reflect the face-to-face interaction between two individuals \cite{eagle2009}, we have seen that the same amount of communication events can correspond to ties with completely different temporal properties, which indicates the necessity to incorporate temporal patterns of human activity in the description and modeling of human interactions. 
Actually, in line with recent work on temporal networks \cite{toivonen2007,holme2012,Karsai:2011is,tang2010} we have shown that the current definition of social ties can be largely improved by taking into account simple temporal tie features, such as the level of burstiness of tie communication or the lifetime of the tie.
\begin{figure}[t]
\begin{center}
\includegraphics[width=1\textwidth,clip=]{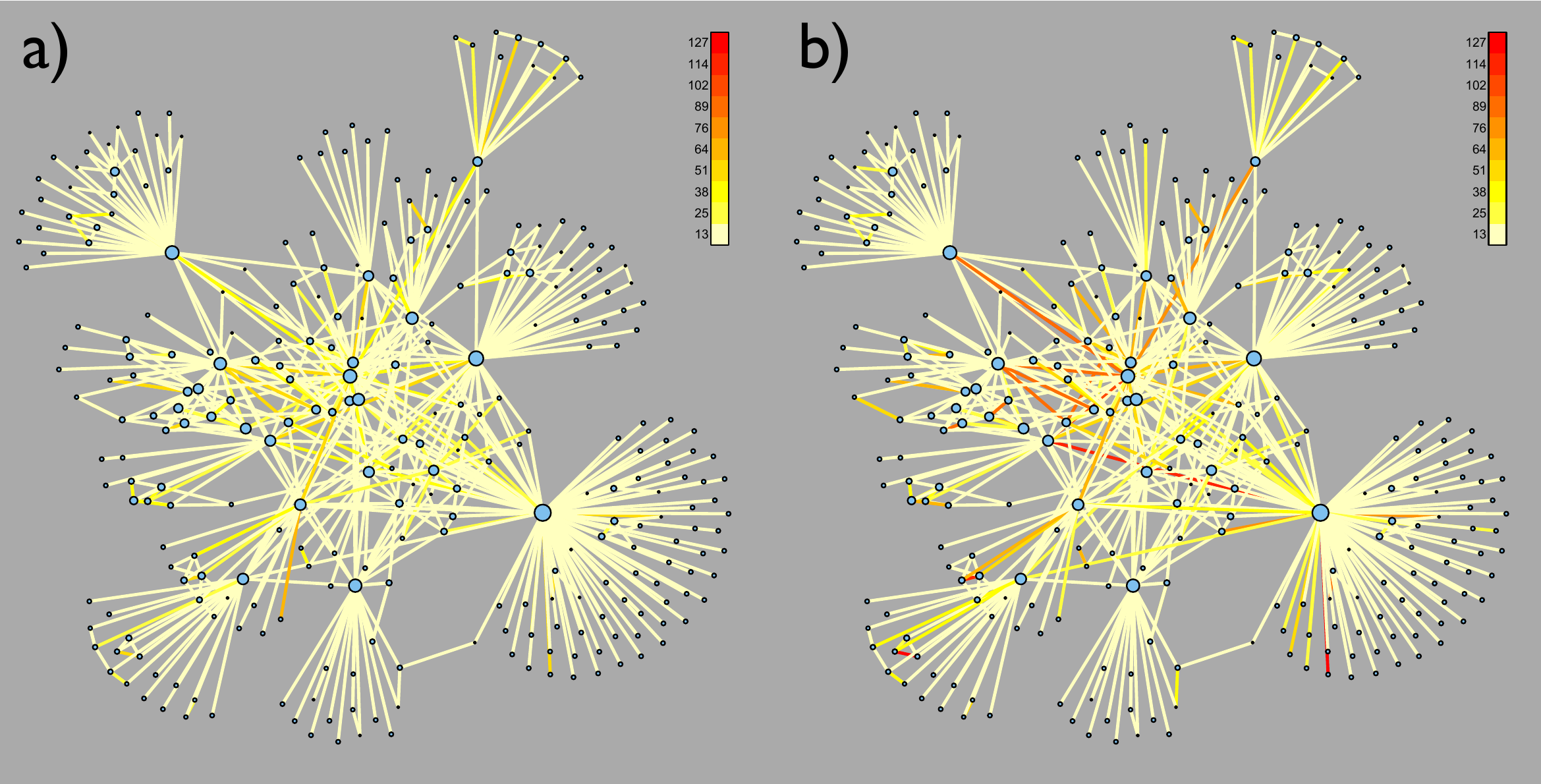}
\end{center}
\caption{(a) Temporal and (b) static structure of the mobile phone network around a randomly chosen individual where colors of the links are proportional to the strength of the ties. The strength of ties is taken as $w_{ij}$ for the static structure, while for the sake of comparison we take the effective weight $w_{ij}(\lambda,T)^* = - T_0/ (\lambda T)  \ln(1-T_{ij}(\lambda,T))$ for the temporal case. Note that in the Poisson case $w_{ij} = w_{ij}(\lambda,T)^*$. Parameters for the calculation are $\lambda=0.5$ and $T=15$ days. It is interesting to note that most of the strong ties present in the static structure become very weak in the temporal picture due to the burstiness and/or lifetime of links.}
\label{fig:grafo}
\end{figure}

Our main contribution is that temporal properties of tie activity also reveal important information on the social topological structure around a given individual. This correlation between human communication and network topology gives important insights on the way in which people distribute their time across their network and, at the same time, it also plays a fundamental role on more global phenomena, such as influence and information spreading in social networks. Previous studies showed that the aggregated time that people dedicate to their connections is affected by cognitive and temporal constraints, which applies in particular to people with larger social circles (hubs) \cite{dunbar1998,goncalves2011,miritello2012b}. Here we have seen that time and/or attention constraints also reflect in the temporal properties of social ties. We observed in fact that, on average, social connections of highly connected individuals not only are weaker in terms of volume of communication, but they are also shorter in time. 
The latter result signals a time allocation strategy of individuals in their social neighborhood, which also affects the transmissibility of a given individual, i.e., her capacity to propagate a piece of information. In particular, we have found that the static picture of networks in terms of social connectivity $k_i$ and/or intensity of communication $s_i$, typically overestimates the power of individuals to transmit information, an effect which is larger for more connected people. 

Our mapping approach goes beyond the important application to information diffusion and it  also applies to the more general and critical problem in complex networks of describing empirical temporal social networks \cite{holme2012}. We have shown, in fact, that a simple way to account for the temporal properties of human communication and still have a static representation of the social network, is by using the transmissibility $T_{ij}$ as a measure of tie strength, instead of the volume of communication (number of calls or total duration) $w_{ij}$. While the number of communication events between a tie $i \to j$ represents the {\it static strength}, the transmissibility $T_{ij}$ represents the {\it dynamical strength} of a tie \cite{miritello2011}. By definition, in fact, $T_{ij}$ incorporates not only the volume of communication $w_{ij}$, but also all the temporal inhomogeneities of human interaction (see Eq.~(\ref{eq:tras2})). The use of the dynamical strength of ties ${\cal T}_{ij}[\lambda,T]$ allows possible comparison of the network at different time scales and thus it can be use to obtain essential information about how and why temporal networks unfold in time. In Fig.~\ref{fig:grafo} we compare a portion of the social network where the weight of each link is given by the static (a) and the dynamical (b) strength of the ties. The figure clearly shows that these two quantities may lead to quite different pictures of the topological structure of the network.  

More generally, this effective static structure of temporal networks might be used in other areas of network research such as determination of influence/centrality \cite{aral2012}, community finding algorithms \cite{fortunato2010}, targeting in viral marketing \cite{iribarren2009}, etc., to analyze the impact of the dynamical patterns of communication in those areas and to assess more accurately the role played by individuals or groups in dynamical process on social networks.

\section{Acknowledgments}
We would like to thank Telef\'onica for providing access to the anonymized data. E.M. and G.M. acknowledge funding from Ministerio de Educaci\'on y Ciencia (Spain) through projects i-Math, FIS2006-01485 (MOSAICO), and FIS2010-22047-C05-04.

\bibliographystyle{acm}
\bibliography{gio.bib}

\end{document}